\documentclass[preprints,article,submit,pdftex,moreauthors]{Definitions/mdpi} 
\usepackage{aasMacro}
\renewcommand{\linenumbers}{}
\firstpage{1} 
\pubvolume{1}
\issuenum{1}
\articlenumber{0}
\pubyear{2023}
\copyrightyear{2023}
\externaleditor{Academic Editor: Dario Gasparrini}
\datereceived{1 December 2022} 
\daterevised{17 January 2023} 
\dateaccepted{19 January 2023} 
\datepublished{} 
\hreflink{https://doi.org/10.3390/galaxies11010022} 
\pdfoutput=1

\Title{Astro-COLIBRI 2---An Advanced Platform for Real-Time Multi-Messenger Discoveries}

\TitleCitation{Astro-COLIBRI 2---An Advanced Platform for Real-Time Multi-Messenger Discoveries}
\TitleCitation{Astro-COLIBRI 2.0}

\Author{P. Reichherzer $^{1,2,3}$*\orcidA{}, F. Schüssler $^{3}$\orcidB{}, V. Lefranc$^{3}$\orcidC{}, J. Becker Tjus$^{1,2}$\orcidD{}, J. Mourier$^{3}$, A.K. Alkan$^{3,4}$\orcidF{}}

\AuthorNames{P. Reichherzer, F. Schüssler, V. Lefranc, J. Becker Tjus, J. Mourier, A.K. Alkan}

\AuthorCitation{Reichherzer, P. et al.}

\address{%
$^{1}$ \quad Ruhr-Universität Bochum, Universitätsstraße 150, D-44801 Bochum, Germany\\
$^{2}$ \quad Ruhr Astroparticle and Plasma Physics Center, Ruhr-Universität Bochum, D-44780 Bochum, Germany\\
$^{3}$ \quad IRFU, CEA, Université Paris-Saclay, F-91191 Gif-sur-Yvette, France\\
$^{4}$ \quad Université Paris-Saclay, CNRS, Laboratoire interdisciplinaire des sciences du numérique, F-91405, Orsay, France}

\corres{Correspondence: astro.colibri@gmail.com}

\abstract{The study of flaring astrophysical events in the multi-messenger approach requires instantaneous follow-up observations to better understand the nature of these events through complementary observational data. We present Astro-COLIBRI as a platform {that integrates} specific tools in the real-time multi-messenger ecosystem.
The Astro-COLIBRI platform bundles and evaluates alerts about transients from various channels. It further automates the coordination of follow-up observations by providing and linking detailed information through its comprehensible graphical user interface. We present the functionalities {with} documented examples of Astro-COLIBRI usage through the community since its public release in August 2021. We highlight the use cases of Astro-COLIBRI for planning follow-up observations by professional and amateur astronomers, as well as checking predictions from theoretical models.}


\begin{document}
\small{\textbf{Disclaimer:} This is the \textbf{Accepted Manuscript} version of an article accepted for publication in \textbf{MDPI Galaxies}. The statements, opinions and data contained in all publications are solely those of the individual author(s) and contributor(s) and not of MDPI and/or the editor(s). MDPI and/or the editor(s) disclaim responsibility for any injury to people or property resulting from any ideas, methods, instructions or products referred to in the content. The Version of Record is available online (doi:  \textbf{10.3390/galaxies11010022}).}

\section{Introduction}
{Advances in real-time multi-messenger observations of the most energetic explosions in the universe have recently~\cite{IceCube:2014stg, Abbott:2017ApJ} contributed significantly to understanding their nature by combining complementary but contemporaneous multi-messenger observations.} Sources of these events exhibit temporal variability on various scales and include gamma-ray bursts (GRBs), {flares of} active galactic nuclei (AGN), supernova (SN) explosions, and~fast radio bursts (FRBs). Advances in the multi-messenger and multi-wavelength approach are driven by next-generation monitoring and follow-up observatories combined with an ecosystem of numerous complimentary services dedicated to a rapid information exchange~\cite{Dorner:2021}.

\begin{figure}[H]
\includegraphics[width=12 cm]{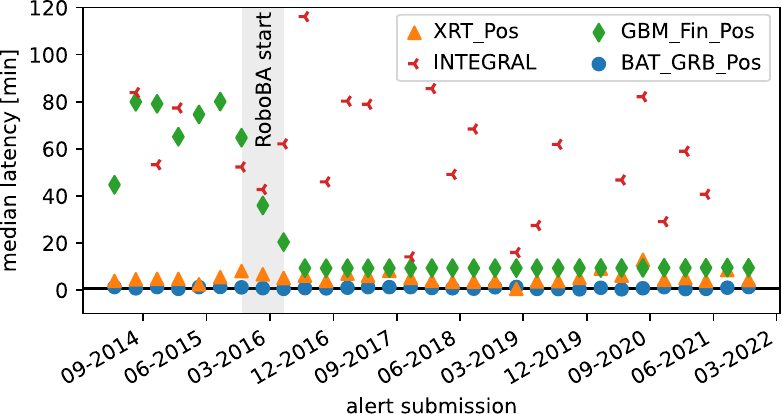}
\caption{The median latencies of alerts per quarter since 2014 for the final VOEvents of the {types} 
 \textit{GBM\_Fin\_Pos}, \textit{INTEGRAL\#Offline}, \textit{XRT\_Pos}, and~\textit{BAT\_GRB\_Pos}. The~vertical bar indicates approximately the start of the automated localization algorithm RoboBA used for the GBM final position alerts. {Taken} 
 from~\cite{Reichherzer:phd:2022}.\label{fig:1}}
\end{figure}

The ensemble of large field-of-view satellites and ground-based observatories around the world constitute the fundamental pillar for multi-messenger astronomy.
They continuously monitor the entire sky for transient behavior across the whole spectrum of detectable messengers. In~the case of the detection of a temporal excess in the real-time analysis that meets telescope-, event-type-, and~messenger-specific criteria, alerts are distributed, and~further offline analyses are triggered. Note that an observatory may distribute multiple alerts for an event at varying stages in the processing, ranging from preliminary event information within the first few seconds to several minutes, hours, or~even days for final analysis results. Interventions may introduce additional latency from humans caused by issues in automated processes. This is illustrated in Figure~\ref{fig:1}, which shows the median latencies \endnote{{Note} 
 that the latency outliers caused by human-in-the-loop interventions due to issues in the automatic pipelines are hidden.} of all alerts per quarter since 2014 for the final GRB VOEvents \endnote{\url{https://voevent.readthedocs.io/en/latest/} (accessed on 23 December 2022)} of INTEGRAL (International Gamma-Ray Astrophysics Laboratory), \emph{{Fermi} 
} GBM {(Gamma-ray Burst Monitor)}, {Swift-BAT} {(Burst Alert Telescope)}, and~XRT {(X-ray Telescope)}. The~use of automatic Fermi RoboBA after 2016 resulted in constant {median} latency of GBM~alerts. 

Thanks to fully automatic processing, {Swift-BAT} and XRT alert latencies are also rather constant. While rapid, automatic alerts are available, the~final offline INTEGRAL alerts, however, show significant variability due to manually generated alerts after interactive data analysis performed by an on-duty~scientist. 

There are several channels for the distribution of notifications of astrophysical transients using different platforms, including {GCN (Gamma-ray Coordinates Network; now 'General Coordinates Network')} notices and circulars, the~transient name server (TNS), and~the Astronomer’s Telegram (ATel). 
We start in Section~\ref{sec:meta} with a discussion of the role of Astro-COLIBRI~\citep{Astro-COLIBRI:2021} {(Coincidence Library for Real-time Inquiry)} as a user-centric platform in the current ecosystem of real-time multi-messenger astronomy that listens to the aforementioned alert streams.
We then continue describing the new features since the public Astro-COLIBRI release in August 2021 in Section~\ref{sec:features}. {We discuss use cases of Astro-COLIBRI} in Section~\ref{sec:use-cases} using the example of several transient events, where Astro-COLIBRI was used by the community of real-time, multi-messenger astronomy.  
{In Section~\ref{sec:outlook}, we describe} future features and connections to other tools, as~well as integration into~observatories.

\begin{figure}[H]
\begin{adjustwidth}{-\extralength}{0cm}
\centering
\includegraphics[width=18cm]{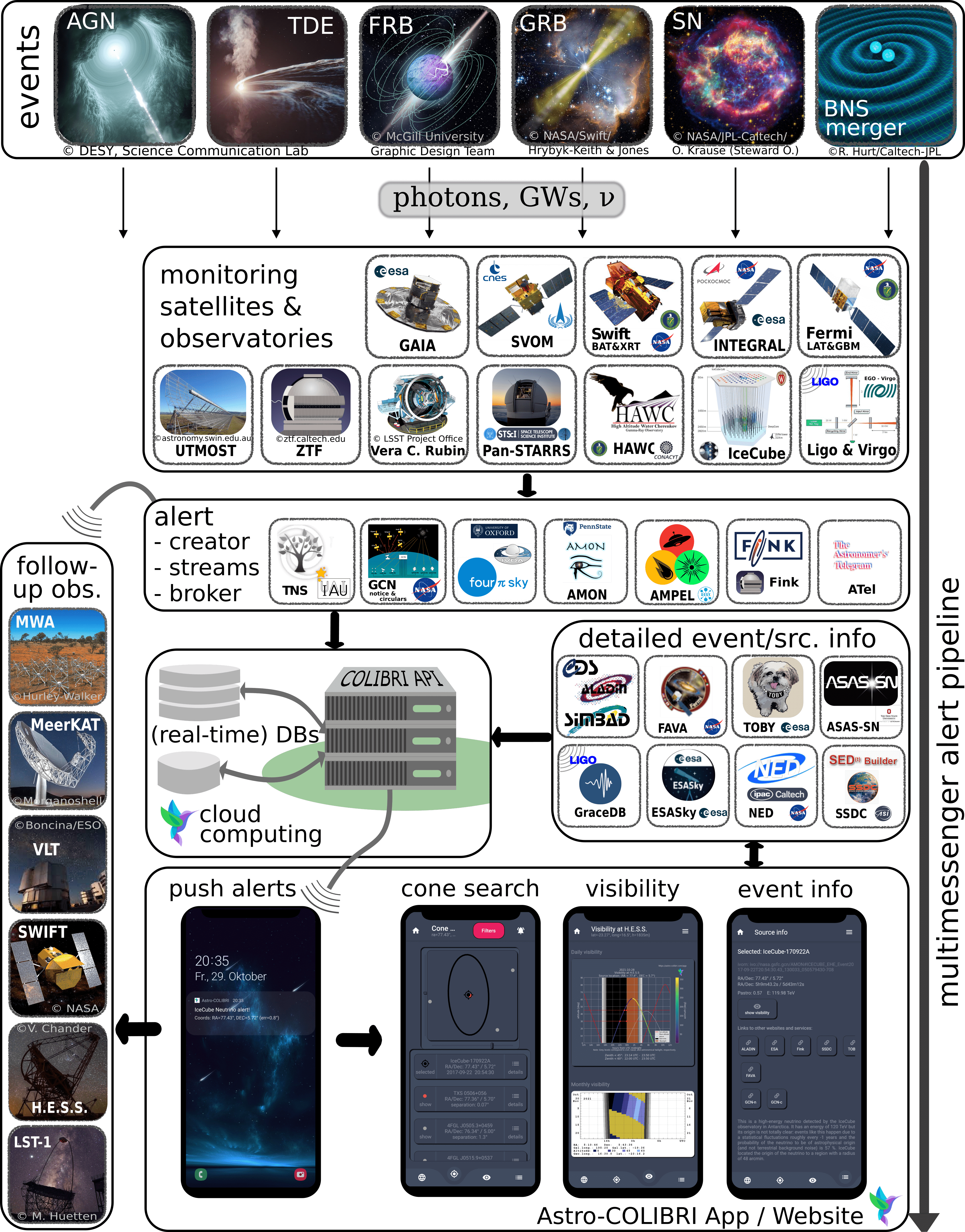}

\end{adjustwidth}
\caption[width=18.3cm]{Overview of the real-time multi-messenger pipeline with boxes summarizing key steps in the workflow to conduct follow-up observations. The~list of shown observatories, tools, and~services is not complete but contains those supported in Astro-COLIBRI. Details are listed in Tables~\ref{tab:alerts} and \ref{tab:3rd_party_services}.\label{fig:2}}
\end{figure}   
\unskip

\section{Astro-COLIBRI as a {Platform}}\label{sec:meta}

\subsection{{Astro-COLIBRI Science~Drivers}}\label{sec:science_drivers}
There are various classes of astrophysical sources that show transient behavior at various messenger types and particle energies. {Figure} 
~\ref{fig:2} shows the transient astrophysical source classes currently supported in Astro-COLIBRI. {In the following list, we show the event types that we display in Astro-COLIBRI. Note that the classes high-energy (HE) neutrinos and gravitational waves (GWs) are messenger types, while all other classes represent astrophysical objects.  The~reason for this is the yet missing real-time and unambiguous association of the two messengers to astrophysical object types.} 

\begin{figure}[H]
    \includegraphics[width=12 cm]{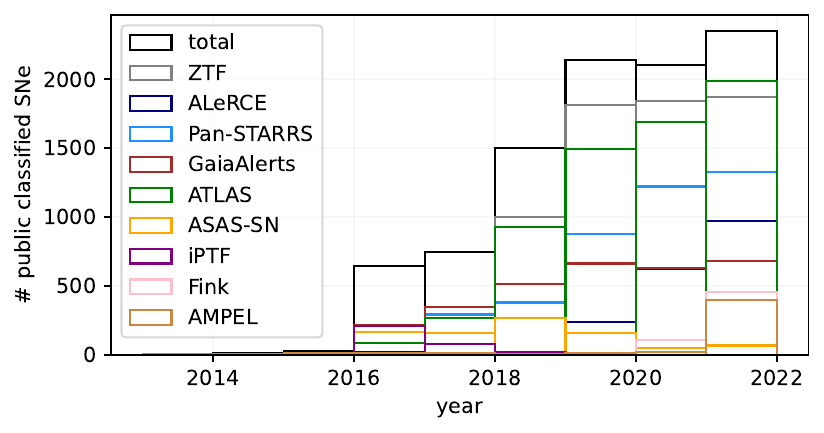}
    \caption{Statistic of publicly distributed and classified {SNe} generated by evaluating data stored in TNS {from} different telescopes or~brokers. \label{fig:sn}}
\end{figure}

\begin{figure}[H]
    \begin{adjustwidth}{-\extralength}{0cm}
    \centering
    \includegraphics[width=17cm]{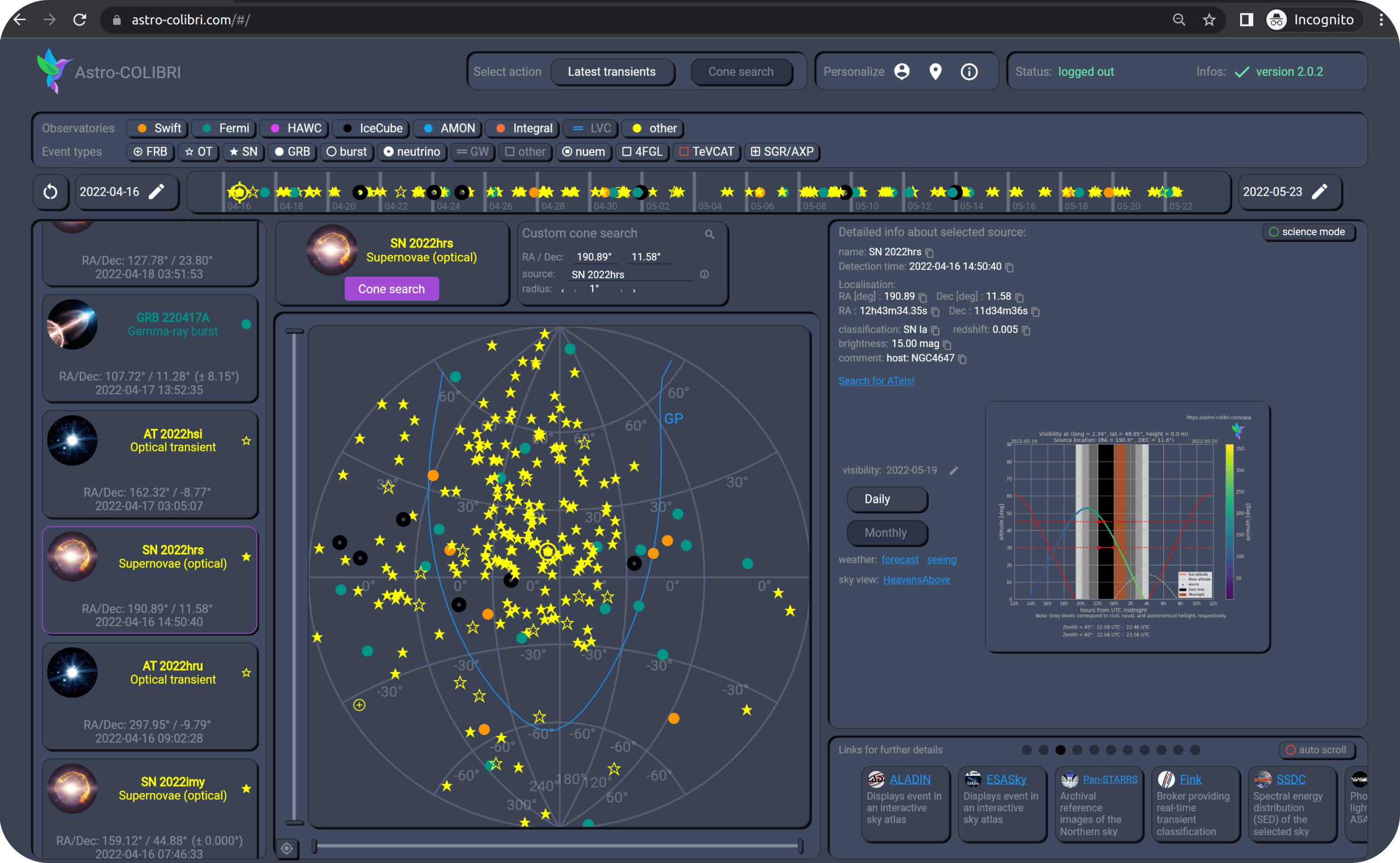}
    
    \end{adjustwidth}
    \caption{{Screenshot} 
 the website of Astro-COLIBRI. In~the top area, settings and filter options are available to customize the interface. The~timeline (see Section~\ref{sec:CS_improved} for details) contains all transient events in the user-defined time range that satisfy the user-set filters. These transients are also shown in the list on the left side and in the sky map. The~right info area displays further information about the selected event/source with customized links to other websites, light curves, the~spectrum, and~at the bottom, detailed (and long-term) visibility information for the specified observatory (see Section~\ref{sec:vis} for details). A~bright example of the SN category is SN {2022hrs}
, which is selected in the screenshot. The~event info provides the brightness (15 mag at detection time). The~visibility in France of the SN for the night of {19 May 2022} 
 is selected to show the time window used for follow-up observations presented in Section~\ref{sec:sns}. \label{fig:transients}}
\end{figure}

\begin{itemize}[leftmargin=10mm,labelsep=3mm,topsep=3pt]
    \item [\textbf{AGN:}]  {There} 
 are different subclasses {of Active Galactic Nuclei (AGN)} with variability on short (minutes/hours) and long (weeks) timescales in $\gamma$-rays, e.g.,~flat-spectrum radio quasars \citep{FermiCatalog:2020}, BL Lac-type objects~\citep{2007ApJ...664L..71A}, and~narrow-line Seyfert 1 objects \citep{Gokus:2021}. There are various mechanisms discussed in literature~\cite{Hawkins:2007, Burd:2021, HESS:2021A&A, Kun:2021} to account for the transient behavior of AGNs, including relativistically accelerated plasmoids moving along the jet axis. Secondary particles are created through various interaction processes that include synchrotron self-Compton, but~also hadronic interaction processes~\cite{Boettcher:2013, Hoerbe:2020}. Transport characteristics of charged particles in these magnetized environments influence the temporal evolution of multi-messenger emission \citep{Reichherzer:2022MNRAS.514.2658R, Reichherzer:2022JOSS, Tjus:2022Physi...4..473B}.
    (Quasi)-periodic {behavior} is observed in supermassive black hole binary systems, caused by, for~example, the~precession of the relativistic jet due to spin-orbit precession~\cite{Kun:2022, Tjus:2022}.
    
    Currently, FLaapLUC \citep{Lenain:2018} alerts on Fermi-LAT detected flares, and~searches for ATels about flaring AGNs are accessible within~Astro-COLIBRI.

    \item [\textbf{GRB:}]   Gamma-ray bursts (GRBs) are powerful explosions typically detected in the \linebreak{X-ray range}. After~the initial short but very intense pulse of high-energy radiation, further photons are emitted subsequently in the afterglow phase, posing an exciting object for multi-wavelength campaigns. 
    Rapid follow-up observations across the electromagnetic spectrum are needed to answer questions about their flare time evolution, the~maximum energy reached, the~progenitor's properties, the~circumburst environment, and~the dependencies of the jet opening angle, etc.
    
    \item [\textbf{FRB:}]   Fast Radio Bursts {(FRBs)} are very brief, ~millisecond long, and~intense bursts of radiation detected in the radio domain. Most of the detected bursts are of extragalactic origin, but~the underlying emission mechanism(s) are still a mystery. Detailed studies in the radio domain are currently being complemented by extensive multi-wavelength campaigns in order to provide crucial insights into these intriguing phenomena. A~major breakthrough was the coincident detection of FRB 200428 and an X-ray burst from the Galactic magnetar SGR 1935+2154 in April 2020~\cite{2020ApJ...898L..29M}. Another such coincidence was reported in October 2022~\cite{2022ATel15708....1L}. 
    Astro-COLIBRI displays FRBs as soon as they are reported to TNS and can be used to illustrate spatial and temporal coincidences with other transients and known emitters, such as SGR 1935+2154. 
    
    \item [\textbf{GW:}] Gravitational waves (GWs) pose an important constituent in the multi-messenger approach as they cannot be absorbed by matter and allow deep insights into source physics. Combined multi-messenger observations that take {GW detections} into account will shed further light on the nature of GRBs, especially when caused by binary neutron star mergers, which are also triggering kilonova~\cite{Abbott:2017ApJ}. Interferometer-based GW observatories, such as LIGO and Virgo, have detected such events over recent  years~\cite{LIGOScientific:2017vwq, LIGOScientific:2021djp, LIGOScientific:2021qlt}. However, detections of current GW observatories come with large uncertainties in sky localization. These large uncertainty regions challenge the search for multi-messenger counterparts and associated sources. Astro-COLIBRI facilitates this task through its graphical interface showing the GW events with their uncertainty region and all temporal and spatial related transients, as well as sources {listed in catalogs, such as 4FGL-DR3 and TevCat}.
    \item [\textbf{HE~$\nu$:}] {The small interaction cross-section of high-energy (HE) neutrinos} allows for deep insights into the central areas of astronomical particle accelerators. The~AGN TXS 0506+056, was the first extragalactic object from which a high-energy neutrino was detected during its flaring state with a temporal and spatial correlation with the significance of $3\sigma$ \cite{IceCube:2018Sci}. 
    Neutrino alerts are displayed in Astro-COLIBRI to help quickly find possible counterparts and assess their multi-wavelength properties and states. Currently, IceCube astrotrack (both gold and bronze){,} as well as cascade-like alerts{,} are~supported.
    
    \item [\textbf{OT:}] Optical transients (OT) is a broad category of phenomena that are typically identified via their transient optical emission. These range from flaring AGNs, tidal disruption events (TDE), cataclysmic variables (CV), stellar flares, novae, fast (blue) transients (FBOT), and~many more.
    
    \item [\textbf{SN:}] Supernovae (SNe) are powerful explosions of massive stars. These transient events can happen in their final life cycle or in a binary system composed of a white dwarf and its companion star. The~number of classified SNe increased over the last years due to improved observatories and more sophisticated analysis pipelines, as~shown in Figure~\ref{fig:sn}, where the number of classified SNe is shown as a function of time broken down to the observatories with the most classified detections. Since version 1.3.0, SNe are supported within Astro-COLIBRI {(see Figure}~\ref{fig:transients})
    , and~users can subscribe to the automatic stream of SNe push~notifications.
    
\end{itemize}

\unskip

However, new transient phenomena are often not clearly categorized. Some alerts may not be associated with a certain emission scenario but rather indicate a special accumulation of individual events in a given observatory. An~example of the latter is the short timescale (0.2 s to 100 s) accumulations of very high-energy photon candidates in the HAWC air shower array. These non-classified events are listed in Astro-COLIBRI within the generic {\it {bursts} 
}category. This category also comprises events that are so rare that they do not justify a dedicated class {within the Astro-COLIBRI interface} on their~own.

\subsection{{Integration of Multi-Messenger~Services}}
{The real-time multi-messenger astronomy community has developed an ecosystem of numerous complementary services and tools over the past decades to study the pressing questions of the transient sky. Those tools, however, come with their own interfaces, pipelines, access points, and~use cases. The~diverse landscape complicates the situation and increases the effort to keep up-to-date with all transient events across the different communities. Here, Astro-COLIBRI acts as a platform that bundles and assesses alerts from these various streams and communities through its comprehensible graphical user interface. Figure~\ref{fig:2} outlines the science drivers, observatories, alert distribution systems (see Table~\ref{tab:alerts} for details), catalogs, and~services currently covered in Astro-COLIBRI, with~details about the services and tools provided in Table~\ref{tab:3rd_party_services}.}

\begin{table}[H]
\caption{Overview of various alert creators, brokers, and~streams. Astro-COLIBRI listens to VOEvent alerts via various brokers and parses notifications from FlaapLUC alert emails and notices from GCN, as~well as from newly confirmed SNe from various alert brokers distributed via~TNS.\label{tab:alerts}}
 \newcolumntype{L}{>{\raggedright\arraybackslash}X}
\begin{tabularx}{\textwidth}{LL}
\toprule
	\multicolumn{1}{c}{\textbf{Service}} & \multicolumn{1}{c}{\textbf{Description}}  \\
	\midrule
	\multicolumn{1}{m{2.8cm}}{AMON~\cite{AMON:2013}} & \multicolumn{1}{p{10cm}}{{AMON (Astrophysical Multimessenger Observatory Network) is a} system searching through streams of sub-threshold events from multi-messenger facilities for correlations of spatial and temporal coincidences.}   \\
	\multicolumn{1}{m{2.8cm}}{AMPEL~\cite{AMPEL:2019}} & \multicolumn{1}{p{10cm}}{Alert management, photometry, and~evaluation of light curves (AMPEL) system. AMPEL combines the functionality of an alert broker with a generic framework to host user-contributed analysis scripts. The~implementation of this stream is currently in the testing phase.}  \\
	\multicolumn{1}{m{2.8cm}}{ATel~\cite{ATels:1998}} & \multicolumn{1}{p{10cm}}{Astronomer’s Telegrams (ATels) are human-written, web-based notifications to report astronomical observations of transient sources.}  \\
	\multicolumn{1}{p{2.8cm}}{{Fink} \cite{FINK:2021}} & \multicolumn{1}{p{10cm}}{Broker for automatized ingestion, annotation, selection, and~redistribution of transient alerts that provides real-time transient classification using deep learning and adaptive learning techniques.} \\
	
	\multicolumn{1}{m{2.8cm}}{FLaapLUC~\cite{Lenain:2018}} & \multicolumn{1}{p{10cm}}{{FLaapLUC (Fermi-LAT automatic aperture photometry Light C$\leftrightarrow$Urve)} detects relative flux variations in \emph{Fermi}-LAT data of sources and alerts users via emails. Based on the Science Tools provided by the \emph{Fermi} Science Support Center and the \emph{Fermi}-LAT collaboration.} \\
	\multicolumn{1}{m{2.8cm}}{Four $\pi$ sky~\cite{4pisky:2016}} & \multicolumn{1}{p{10cm}}{Infrastructure of open data-services based on the VOEvent standardized message-format for distributing transient alerts, such as a VOEvent broker and a DB that stores historical transient alerts.}\\
	\multicolumn{1}{m{2.8cm}}{GCN-\newline Circulars~\cite{GCNs:2000}} & \multicolumn{1}{p{10cm}}{The Gamma-ray Coordinates Network (GCN, now 'General Coordinates Network') distributes human-written reports (named Circulars) about follow-up observations of GRBs.} \\
	\multicolumn{1}{m{2.8cm}}{GCN-Notices~\cite{GCNs:2000}} & \multicolumn{1}{p{10cm}}{Machine-generated alerts about new detections intended to trigger rapid follow-up campaigns.} \\
	\multicolumn{1}{m{2.8cm}}{TNS \textsuperscript{a}} & \multicolumn{1}{p{10cm}}{The Transient Name Server (TNS) is the official International Astronomical Union mechanism for reporting new SN candidates and naming spectroscopically confirmed ones. There are dedicated brokers developed to handle large-scale astronomical survey data from ZTF and LSST e.g.,~such as AMPEL~\cite{AMPEL:2019}, {Fink}~\cite{FINK:2021}, ANTARES~\cite{ANTARES:2014}, ALeRCE~\cite{ALeRCE:2021}, Lasair~\cite{Lasair:2019}, MARS \textsuperscript{b}, and~Pitt-Google Broker \textsuperscript{c} that report detections and classifications via TNS.}  \\
	\multicolumn{1}{m{2.8cm}}{VOEvents~\cite{VOEvents:2017}\newline} & \multicolumn{1}{p{10cm}}{In order to use a standardized, machine-readable format, VOEvents {(VO stands for Virtual Observatory)} were officially adopted to report transients in 2006 by the International Virtual Observatory Alliance {(IVOA)}.}  
	\\
	\bottomrule
\end{tabularx}

\footnotesize{\textsuperscript{{a} 
}  \url{https://www.wis-tns.org/};
\textsuperscript{b}  \url{https://mars.lco.global/};
\textsuperscript{c}  \url{ https://pitt-broker.readthedocs.io/en/latest/}.}
\end{table}
\unskip

\begin{table}[H]
\caption{Astro-COLIBRI is embedded in the ecosystem of real-time multi-messenger services. Customized links point directly from Astro-COLIBRI to the services and websites listed in the table. Figure~\ref{fig:transients} {shows} 
 in the bottom right area the list of customized links for each event and source to provide convenient access to specific, more detailed information on~demand.\label{tab:3rd_party_services}}
 \newcolumntype{L}{>{\raggedright\arraybackslash}X}
\begin{tabularx}{\textwidth}{LL}
\toprule
	\multicolumn{1}{c}{\textbf{Service}} & \multicolumn{1}{c}{\textbf{Description}}  \\
	\midrule
		\multicolumn{1}{p{2.6cm}}{ASAS-SN \textsuperscript{e}} & {All Sky Automated Survey for SuperNovae (ASAS-SN) is an automatic all-sky survey to detect SNe. The~link allows to obtain a photometric light curve of the selected sky region.} \\
		\multicolumn{1}{p{2.6cm}}{ALADIN~\cite{Aladin:2014}} & {Displays sources from {catalogs} or databases (DBs) in a sky map.}   \\
            \multicolumn{1}{p{2.6cm}}{ALeRCE~\cite{ALeRCE:2021}} & {Web portal for transients from ALeRCE (Automatic Learning for the Rapid Classification of Events).}   \\
		\multicolumn{1}{p{2.6cm}}{BAT~\cite{BAT:2013}} & {Burst Alert Telescope (BAT) GRB event data processing report including detailed info, a~flux summary, spectra, and~light curves.}  \\
		\multicolumn{1}{p{2.6cm}}{ESA~\cite{ESASKY:2018, ESA:2022}} & {ESASky is an application that visualizes astronomical data in a sky map.}  \\
		\multicolumn{1}{p{2.6cm}}{FAVA~\cite{FAVA:2017}} & {\emph{Fermi} All-sky
        Variability Analysis (FAVA) for the selected position observed by the Large Area Telescope (LAT).}  \\
        \multicolumn{1}{p{2.6cm}}{{Fink} \cite{FINK:2021}} & {Broker and science portal providing information about transients.} \\
        \multicolumn{1}{p{2.6cm}}{Gaia~\cite{Gaia:2021}} & {Gaia Photometric Science Alerts is an all-sky photometric transient survey, based on the repeated measurements of Gaia.}  \\
		\multicolumn{1}{p{2.6cm}}{GBM \textsuperscript{f}} & {Quicklook directory with near real-time (10--15 min after the \emph{Fermi}-{Gamma-ray Burst Monitor} (GBM) trigger) information of light curves and spacecraft pointing history. } \\
		\multicolumn{1}{p{2.6cm}}{GCN-c \textsuperscript{g}} & {Gamma-ray Coordinates Network circulars (GCN-c) inform in human-written reports about the observed event.}  \\
		\multicolumn{1}{p{2.6cm}}{GCN-n \textsuperscript{h}} & {GCN notices (GCN-n) deliver information about basic properties of transient objects in automatically generated alerts.} \\
		\multicolumn{1}{p{2.6cm}}{GraceDB \textsuperscript{i}} & {Communications hub and DB with information about candidate gravitational-wave events.} \\
		\multicolumn{1}{p{2.6cm}}{IBAS~\cite{IBAS:2013}} & {System for real-time detection of GRBs seen by INTEGRAL.} \\
		\multicolumn{1}{p{2.6cm}}{LAT-LCR \textsuperscript{j}} & {DB of multi-cadence flux calibrated light curves for over 1500 variable sources from the 10 year \emph{Fermi}-LAT point source catalog.} \\
		\multicolumn{1}{p{2.6cm}}{NED~\cite{NED:1991}} &  {NASA/IPAC Extragalactic Database (NED)} about galaxies and other {extragalactic} objects, with~more details, photometry, spectra and further references/links. \\
		\multicolumn{1}{p{2.6cm}}{Pan-STARRS~\cite{PAN_STARRS:2020}} & {DB of images after analysis and processing obtained by the Pan-STARRS {(Panoramic Survey Telescope And Rapid Response System)} telescopes.} \\
		\multicolumn{1}{p{2.6cm}}{SkyMapper~\cite{SkyMapper:2019}} & {DB with a digital record of the entire southern sky, which stores images and catalogs from SkyMapper's Southern Survey.}  \\
		\multicolumn{1}{p{2.6cm}}{SIMBAD~\cite{Simbad:2000}} &  {Provides additional information, cross-identifications, bibliography and measurements of sources.}  \\
            \multicolumn{1}{p{2.6cm}}{SNAD~\cite{Snad:2022}} &  {Web portal for objects from the Zwicky Transient Facility’s.}  \\
		\multicolumn{1}{p{2.6cm}}{SSDC \textsuperscript{k}} & {Display of Spectral Energy Distributions (SEDs) of astrophysical sources from the Space Science Data Center (SSDC).} \\
		\multicolumn{1}{p{2.6cm}}{Swift \textsuperscript{l}} & {High-level information on the {Swift-BAT} or XRT observations and links to other relevant references.}  \\
		\multicolumn{1}{p{2.6cm}}{TACH~\cite{TACH:2020}} & {The Time-domain Astronomy Coordination Hub (TACH) is a NASA GSFC project that provides an overview over GCN notices and circulars via the ``GCN viewer'' interface. }   \\
		\multicolumn{1}{p{2.6cm}}{TeVCat~\cite{TevCat:2008}} & {Source catalog for very high energy gamma-ray astronomy with detailed source information and linked references.}   \\
		\multicolumn{1}{p{2.6cm}}{TNS \textsuperscript{m}} & {The Transient Name Server (TNS) provides discovery and classification reports of e.g.,~SNe and further info, e.g.,~spectra.}   \\
		\multicolumn{1}{p{2.6cm}}{TOBY \textsuperscript{n}} & {The Tool for Observation visiBilitY and schedule (TOBY) shows event visibility and schedule for a variety of observatories.} \\
		\multicolumn{1}{p{2.6cm}}{XRT~\cite{XRT:2009}} & {Light curve, spectra, and~comparison with other bursts provided by Swift's X-ray Telescope (XRT).} \\
			\bottomrule
\end{tabularx}
{\footnotesize{
All links accessed on 23 December 2022; 
\textsuperscript{{e} 
} \url{https://www.astronomy.ohio-state.edu/asassn/index.shtml}; 
\textsuperscript{f} \url{https://heasarc.gsfc.nasa.gov/FTP/fermi/data/gbm/triggers/};
\textsuperscript{g} \url{https://gcn.gsfc.nasa.gov/gcn/gcn3_archive.html};
\textsuperscript{h} \url{https://gcn.gsfc.nasa.gov/};
\textsuperscript{i} \url{https://gracedb.ligo.org/latest/};
\textsuperscript{j} \url{https://fermi.gsfc.nasa.gov/ssc/data/access/lat/LightCurveRepository/};
\textsuperscript{k} \url{https://tools.ssdc.asi.it/SED/};
\textsuperscript{l} \url{https://swift.gsfc.nasa.gov/archive/grb_table/};
\textsuperscript{m} \url{https://www.wis-tns.org/};
\textsuperscript{n} \url{http://integral.esa.int/toby/}
}}\\
\end{table}
\unskip

\subsubsection{{Alert Distribution~Systems}}
Alerts of transient events and assessing their relevance for follow-up observations to the respective observatories are of fundamental importance. Rapid distribution of these events occurs via several channels on different platforms and timescales. Alerts streams refer to the sequence of distributed~alerts.

The various alert systems used for this task are summarized in Table~\ref{tab:alerts}, including VOEvents, GCNs (notices and circulars), TNS notifications, and~ATels. 
These alert distribution mechanisms differ in their underlying communication technology, the~targeted scientific community, the~degree of machine- or human-generated content of the alerts, and~the latencies. The~applied distribution technologies involve the usage of application programming interfaces, brokers, databases, emails, mobile applications, and~webpages. The~patchwork of different mechanisms requires the community to make an effort to keep track of the ever-increasing amount of relevant~information.

\begin{figure}[H]
\includegraphics[width=12.0 cm]{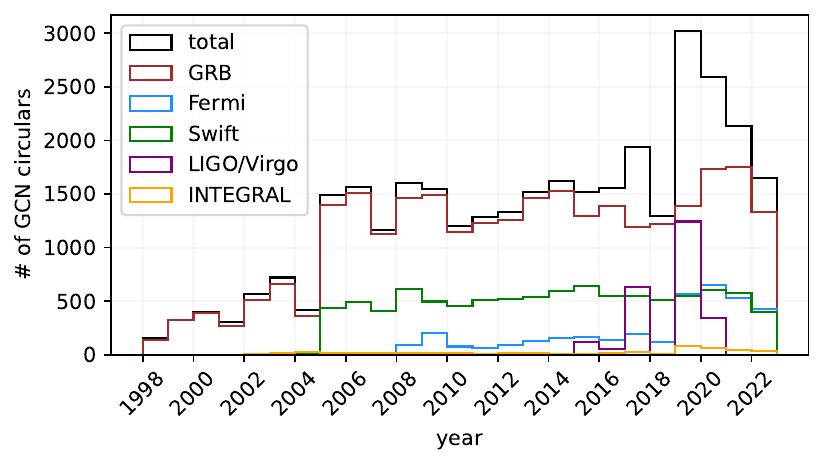}
\caption{Number of GCNs circulars per year. The~number has steadily increased over the past few years. A~strong increase was due to the addition of GW~observations.\label{fig:gcns}}
\end{figure}

For example, the~dominant pipeline for GRB transients starts with the monitoring observatories \emph{Fermi}, Swift, or~INTEGRAL that send their detections via the Gamma-ray Coordinates Network (GCN, also standing for General Coordinates Network) as notices or circulars to the community. Detections are typically announced via the largely automatized and machine-readable GCN notices. Human-written GCN circulars are then used to summarize and confirm the detections. The~statistics of GCN circulars are shown in Figure~\ref{fig:gcns}{, where it can be seen} that alerting GCN circulars from these monitoring observatories make up only a tiny fraction, as~most GCN circulars report follow-up observations from ground-based observatories around the globe, covering a broad range of wavelengths. Alerts of optical transients are, however, predominantly distributed through other pipelines and streams, including, e.g.,~TNS or brokers, such as AMPEL and {Fink}.

\unskip   

\subsubsection{{Event Information~Databases}}
{In recent decades, an~ecosystem of services was created to study the various types of different transients. This ecosystem provides specific information on various aspects relevant to real-time multi-messenger astrophysics.} These services distribute information via static or dynamic web pages, databases, queryable application programming interfaces (APIs), or~interactive sky maps. {We summarize in Table~\ref{tab:3rd_party_services}} the systems linked within Astro-COLIBRI, each offering a particular value for studying transient events. Note that most services focus on specific types of events and only provide a comprehensive view of transient events when combined with other services in the~table.

\section{{New~Features} of Astro-COLIBRI}\label{sec:features}
Astro-COLIBRI has changed significantly since its release in August 2021. User input has been leveraged to include additional features, event categories, notification streams, and~use cases. The~interfaces have also become more intuitive and informative over the last 12 updates of Astro-COLIBRI \endnote{{The}
 changelog of all versions are documented on \url{https://astro-colibri.science/documentation}  (accessed on 23 December 2022).}.
We briefly overview new functionalities in the {following}, leaving the full details to the documentation, as~they are liable to change as the platform is improved further:
\unskip
\subsection{Timeline}
The novel timeline feature complements the existing display of transient events to visualize temporal correlations. This feature is located below the filter area and between both date buttons (see Figure~\ref{fig:transients}). The~interactive timeline also facilitates the choice of the time period through clickable vertical lines and the surrounding buttons. The~timeline also works in the cone-search view, presenting a high-level temporal overview of transient events in a given sky~region.
\subsection{Improved Cone~Searches}\label{sec:CS_improved}
Cone searches are one of the main functions of Astro-COLIBRI. This presents a transient event in the context of other events or sources in the relevant temporal and spatial phase space. The~improved cone search of Astro-COLIBRI now also displays the localization uncertainties of all transients in the cone-search view to notice overlapping detections (see the middle pane of Figure~\ref{fig:grb20221009} for an example), including the newly supported transient classes such as SNe. The~sources that are also shown in the cone search now also contain sources from catalogs of soft gamma-ray repeaters (SGRs) and {anomalous} X-ray pulsars (AXPs). Existing catalogs were updated to 4FGL-DR3 and the latest~TevCat.

\subsection{Search for~ATels}\label{sec:ATels}
In particular, for~AGN flares, follow-up observations are reported via ATels. For~all events and sources in Astro-COLIBRI there is now the possibility to search for ATels that can provide reports on historical flares. Currently, ATels in Astro-COLIBRI are accessed via the NASA ADS API. However, the~ATels are submitted to NASA ADS with a delay of one week, which is why the latest ATels are currently missing. This will be fixed in a future version of Astro-COLIBRI {by parsing the ATels distributed by mail and storing the relevant information in our databases.}
\subsection{Novel Notification~Streams}
Since the initial release of Astro-COLIBRI, new transient source classes were added. The~main ones are optical transients (SNe, novae, CVs, etc.) and GeV flares of AGNs detected by Fermi-LAT via the FLaapLUC pipeline. Users can subscribe to these new notification streams within the Astro-COLIBRI app. A~dedicated stream for amateur astronomers was also added that only sends notifications for bright (mag < 18) optical~transients. 

\subsection{Science~Mode}
In order to satisfy the various user profiles, we have implemented a switch between general, easy-to-understand information about events and detailed information for professional planning of follow-up observations. The~\emph{{science} 
 mode} provides details useful for professional campaigns for follow-up observation, such as the visibility plots. When the \emph{science mode} is switched off, general information about the transient event is~provided.

\subsection{Visibility~Plots}\label{sec:vis}
The selection of custom dates for the daily visibility plots is now possible, supporting the investigation of {historical} and future observation conditions. 
We also added the possibility to define and manage custom observatories used for the visibility assessment. In~addition, various new links to external services now provide weather information, seeing estimates and a real-time sky~view.
\begin{figure}[H]
\begin{adjustwidth}{-\extralength}{0cm}
\centering
\includegraphics[width=18cm]{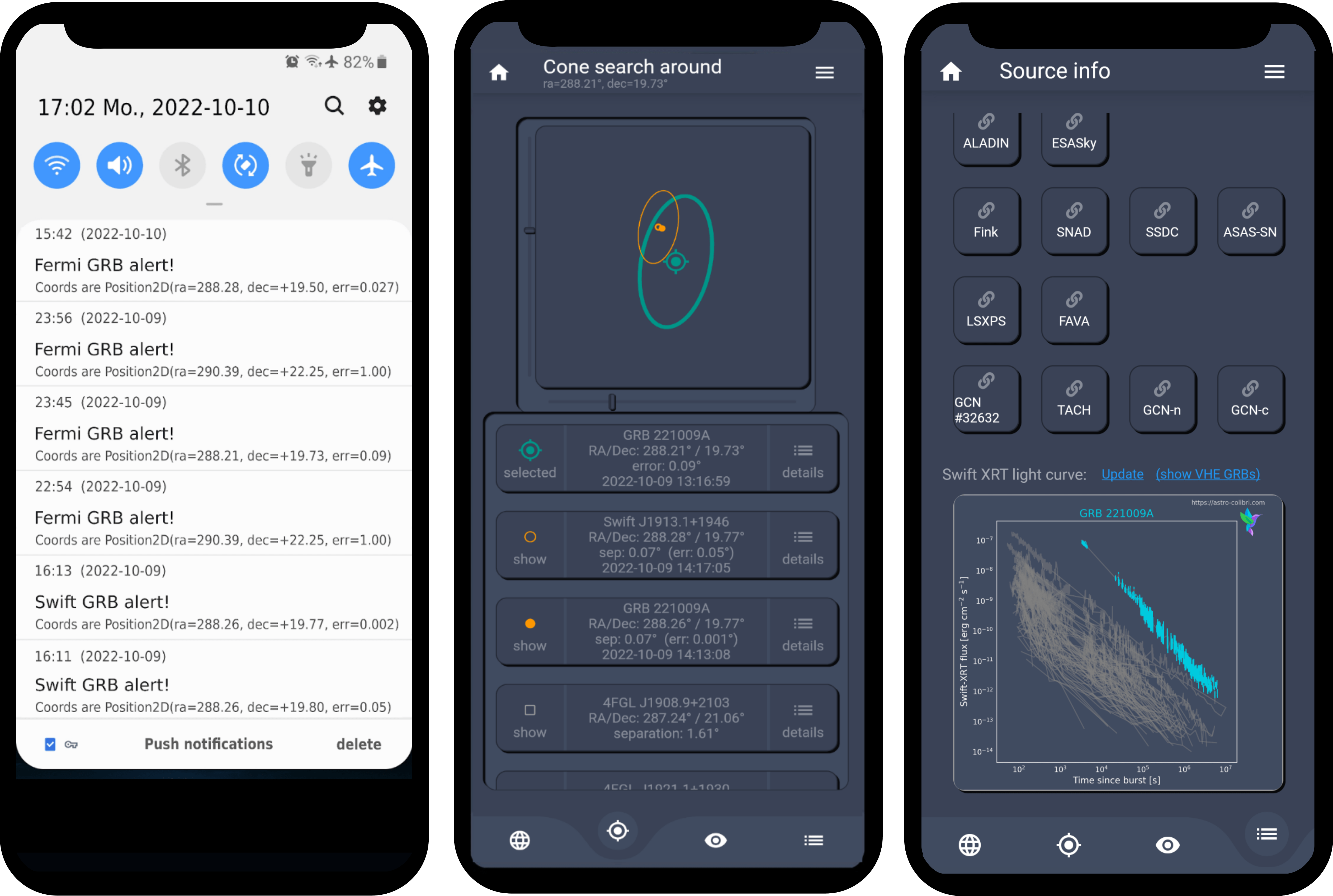}
\end{adjustwidth}
\caption{{Mobile} 
 screens showing push notifications sent via Astro-COLIBRI about GRB 20221009A (see Section~\ref{sec:follow-up}), cone search (see Section~\ref{sec:CS_improved}) illustrating the overlapping Swift and Fermi localizations, and~customized links and the Swift XRT light curve (see Section~\ref{sec:XRT}) in the right screen.\label{fig:grb20221009}}
\end{figure}
\unskip 

\subsection{GRB Light~Curves}\label{sec:XRT}
A new feature in Astro-COLIBRI is the display of the light curve of Swift-XRT GRBs, where the flux is displayed as a function of time since the detection of the burst. For~better evaluation, the~light curve is shown compared to historical GRBs. The~right panel of Figure~\ref{fig:grb20221009} shows the light curve of GRB 20221009A compared to all historical Swift-XRT GRBs. The~enormous brightness of the event becomes immediately clear in this representation. In~general, the~plot also helps estimate fluxes and can facilitate planning of follow-up~observations. 

\subsection{Coordinate Systems and Projections}
Two celestial coordinate systems, namely Galactic and Equatorial, are now implemented in Astro-COLIBRI to specify the positions of celestial objects. The~right-handed convention is used, meaning that the coordinates in the fundamental plane are positive to the north and east. In~Astro-COLIBRI, three different projection systems are available to project astronomical coordinates on the two-dimensional map shown in Figure~\ref{fig:transients}. Each projection method optimizes the display of the celestial objects on the map regarding different properties, while distortions to a certain degree are inevitable. The~Mollweide and the Hammer-Aitov projection are equal-area map projections. The~distortions at large declinations are not as substantial as in the similar-looking Mollweide projection. The~last supported method in Astro-COLIBRI, the~Mercator projection, flattens the spherical sky into a two-dimensional, rectangular map with latitude and longitude lines drawn in a grid. Users can select their favorite combination of coordinate system and projection within the Astro-COLIBRI~platform.

\section{Astro-COLIBRI in~Practice}\label{sec:use-cases}
\unskip

\subsection{Multi-Wavelength Follow-Up~Observations}\label{sec:follow-up}
Astro-COLIBRI allows real-time (mobile) tracking of all relevant updates on exciting transients coupled with tools useful for planning follow-up observations. However, the~functionality of Astro-COLIBRI goes beyond this, as~multi-wavelength follow-up observations reported in ATels can be displayed via the interface using the function described in Section~\ref{sec:ATels}. In~addition, all circulars published via GCN are made available via the link to TACH. In~the following, details are described using GRB 20221009A as an~example:
\begin{itemize}
    \item \textbf{{Gamma} 
-Ray Burst GRB 20221009A} 
    Swift-BAT first reported the historical event GRB 20221009A to the community on {9}
 October 2022 at 14:11:33 UT with a GCN notice (trigger number 1126853\endnote{GCN notices: \url{https://gcn.gsfc.nasa.gov/other/1126853.swift} (accessed on 23 December 2022)}). Two minutes later, Swift-XRT sent its first notice about this event with updated localization information.
    Note that Swift-BAT triggered on the event a second time with the number 1,126,854  \endnote{GCN notices: \url{https://gcn.gsfc.nasa.gov/other/1126854.swift} (accessed on 23 December 2022)} and assigned the name Swift J1913.1+1946 stating that it is a Galactic transient due to the proximity to the galactic plane, the~exceptional brightness and a match with a source from the Swift-BAT on-board catalog. The~Fermi satellite detected the object almost one hour earlier on 2022-10-09 at 13:16:59 UT, but~managed to distribute the alert only eight hours later through a GCN notice due to the failure of their automatic alert pipeline (note that the median latencies of Fermi GBM final position GCNs are within 10 minutes as shown in Figure~\ref{fig:1}). 
    
    The detailed history of the alerts of Fermi and Swift that were distributed through Astro-COLIBRI is shown in Figure~\ref{fig:grb20221009} on the left mobile screen. Here, the~overview of push notifications is shown on 10 October 2022 at 13:02:12 UT, including further GCN circulars with the final localization~information. 
    
    In the middle mobile screen, the~cone search around the GRB 20221009A position of Fermi is presented, including the Swift-XRT (trigger number 1,126,853) and Swift-BAT (trigger number 1,126,854) detections taken on 9 October 2022 at 21:50:02 UT. Note that later Fermi updates modified the localization shown in Astro-COLIBRI~slightly. 
    
    In the right mobile screen in the lower half, the~new feature (see Section~\ref{sec:XRT}) of the Swift-XRT GRB light curves is shown, where the Swift-XRT flux is shown as a function of the time since the burst. Here, the~GRB 20221009A light curve in cyan is compared to historical GRBs. The~flux of GRB 20221009A is significantly larger compared to other historic GRBs. The~ATels search functionality within Astro-COLIBRI provides a long list of reported follow-up observations. The~customized link to TACH showing this event, provides further reports about follow-up multi-wavelength observations.
\end{itemize}

\subsection{Optical Follow-Up Observations of~SNe}\label{sec:sns}
Astro-COLIBRI has become an integral part of many amateur astronomers' quest to study bright supernovae. Users can choose to be informed only about the brightest SNe by push notifications through the app. In~the following, two documented cases are listed where these alerts have led to follow-up observations by amateur~astronomers:
\begin{itemize}
    \item \textbf{Supernova SN 2022eyj} This SN Ia with redshift 0.021 was detected by ASAS-SN on 22 March 2022 07:26:24 UT (ra: 169.50°, dec: 7.85°) with a brightness of 16.1 mag. The~user @SacHA(P) was notified about the SN via Astro-COLIBRI push notifications and performed subsequent optical observations that were reported on Twitter \endnote{\url{https://twitter.com/P_AHcas/status/1506790945672548357?cxt=HHwWisCyubeYmekpAAAA} (accessed on 23 December 2022)}.
    \item \textbf{Supernova SN 2022hrs} This bright (15 mag on detection 16 April 2022 14:50:40) SN Ia with redshift 0.005 is shown as the selected event in Figure~\ref{fig:transients} (ra: 190.89°, dec: 11.58°). The~user @Stef\_Astro was notified about the SN via Astro-COLIBRI push notifications and performed subsequent optical observations that were reported on Twitter \endnote{{\url{https://astro-colibri.science/usecases} (accessed on 23 December 2022). Note that the @Stef\_Astro Twitter account is deleted.}}.
\end{itemize}

We use Twitter for public exchange and sharing of user-performed \linebreak follow-up observations.

\begin{figure}[H]
\includegraphics[width=13cm]{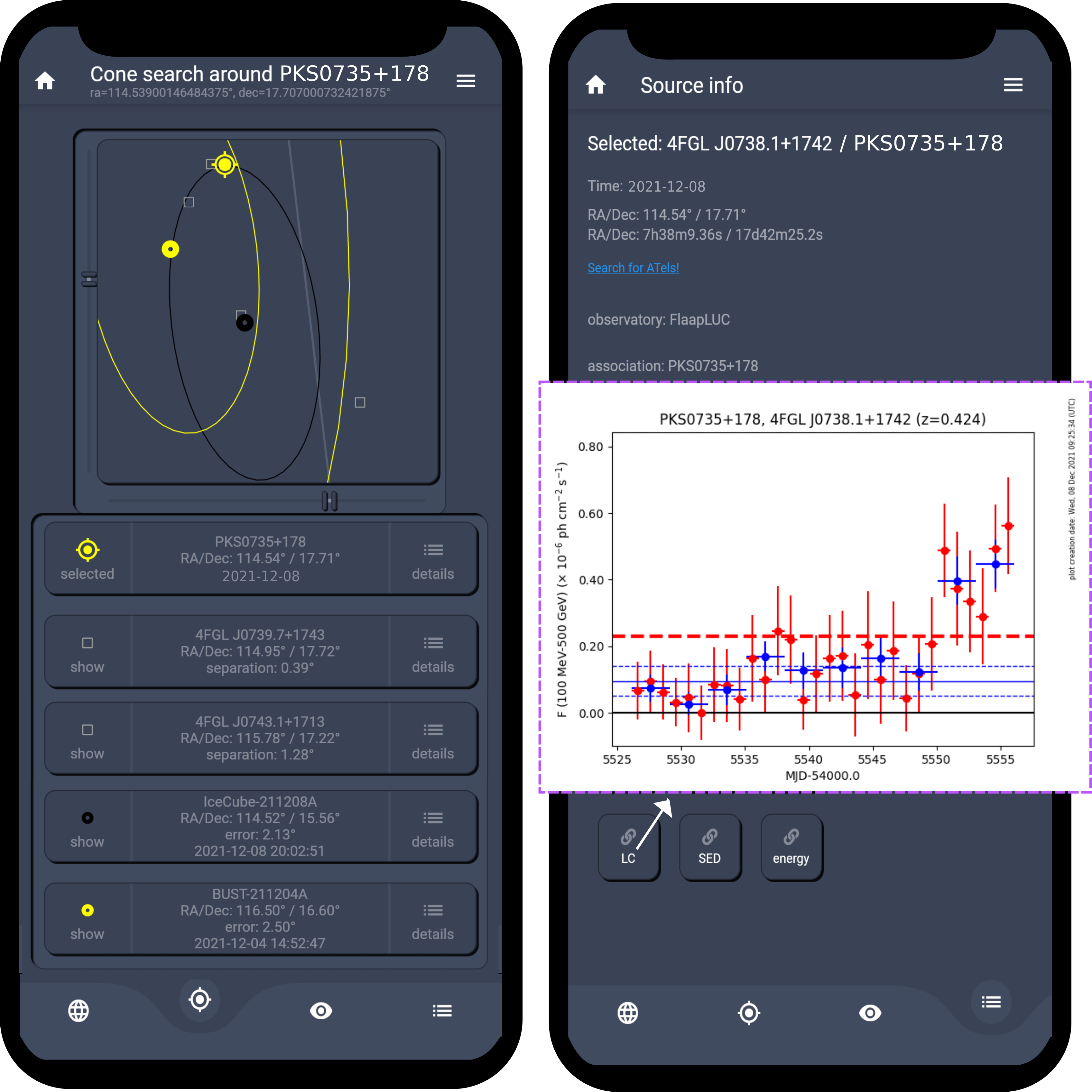}
\caption{{Screenshots} 
 the Astro-COLIBRI app. The~left mobile screen shows the cone search of the FLaapLUC alert from the flaring AGN PKS 0735+178 contained within the uncertainty region (black ellipse) of the neutrino IceCube-211208A (see Section~\ref{sec:CS_improved}). The~list of sources/events in the cone search is shown below, ordered by the separation to PKS 0735+178. The~right mobile screen shows information on PKS 0735+178. Clicking on the light curve (LC) button opens the FLaapLUC light curve plot in the browser.  \label{fig:pks}}
\end{figure}   
\unskip
\subsection{Multiple Alerts from the Same Phase~Space}
With Astro-COLIBRI, it is possible to check whether there are potential connections between multi-messenger detections over arbitrary time periods in a certain phase~space.
\begin{itemize}
    \item \textbf{PKS 0735+178 and IceCube-211208A} A cone search around PKS 0735+178, for~which FLaapLUC issued an alert to report about a flare on {8 December 2022} \endnote{See also \url{https://www.astronomerstelegram.org/?read=15099} (accessed on 23 December 2022)}, reveals that the source position is contained within the uncertainty region of the neutrino event IceCube-211208A reported on the same date{, as can be seen in Figure} \ref{fig:pks}.
\end{itemize}

\subsection{Alerts to Test Theoretical Model~Predictions}
\begin{figure}[H]
        \includegraphics[width=13cm]{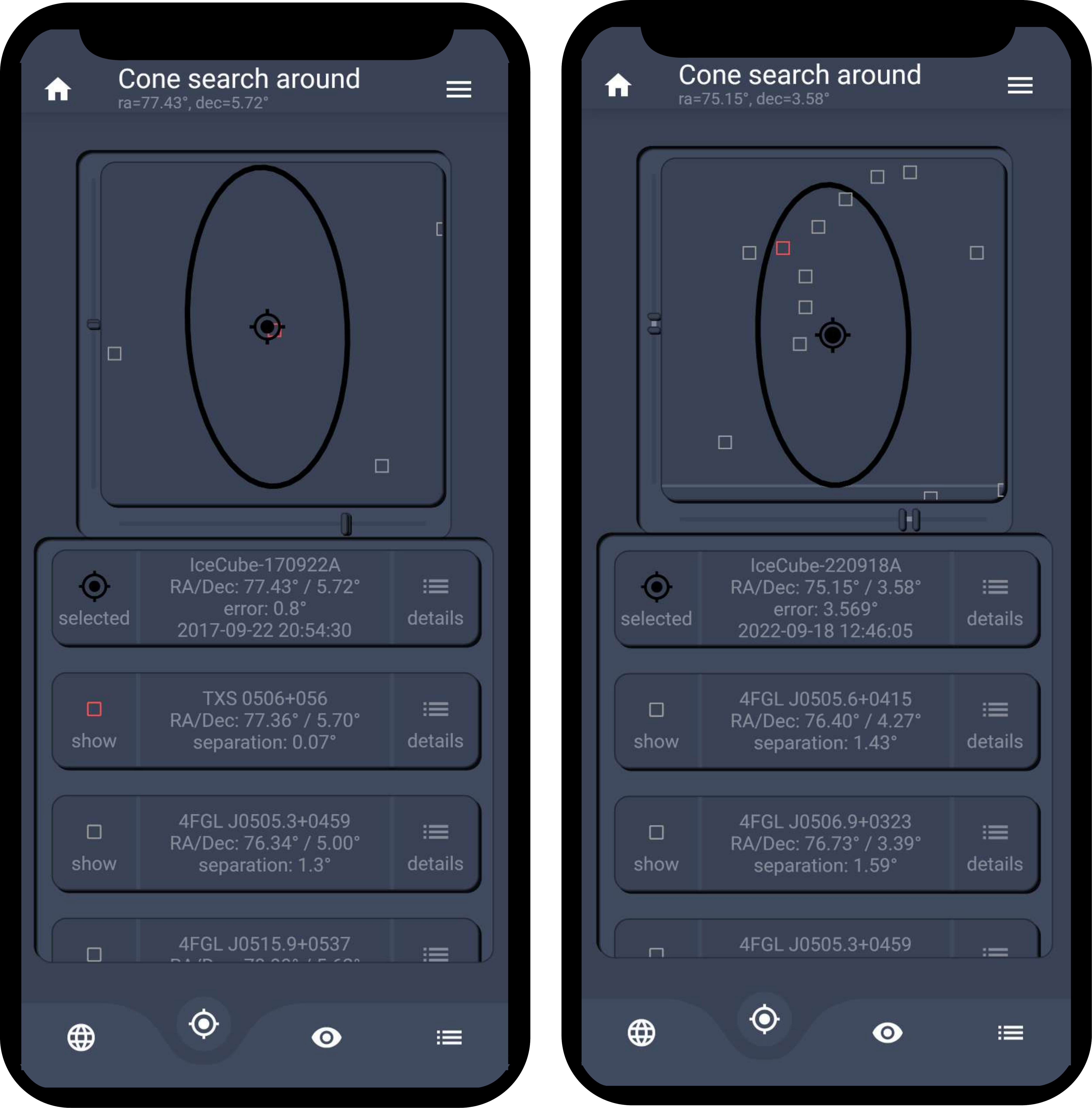}
        \caption{{Mobile} 
 screen of cone searches within the Astro-COLIBRI event for the IceCube$-$170922A (\textbf{left}) and IceCube$-$220918A (\textbf{right}) alerts with their uncertainty regions surrounded by black lines. \label{fig:icecube_TXS}}
    \end{figure}

Theoretical models of transients often predict the emission of multiple messenger types, e.g.,~hadronic jet models predict photon and neutrino emission. Astro-COLIBRI can be employed to find possible counterparts of these emissions and test theoretical model predictions. A~particular use case concerns IceCube neutrino alerts, as~the origin of each high-energy neutrino is unknown during the time of observation. With~Astro-COLIBRI, it is possible to rapidly check for, e.g.,~flaring AGNs within the localization uncertainty region (black ellipses) of the neutrino~detection.   
\begin{itemize}
    \item \textbf{IceCube-170922A and TXS 0506+056} 
    The extended delay in discovering the spatial and temporal correlation between the neutrino event and the AGN flare was driving the development of the Astro-COLIBRI platform. Figure~\ref{fig:icecube_TXS} in the left panel shows the event in the Astro-COLIBRI app in cone-search view, where the black ellipse indicates the localization uncertainty. By~clicking on details and the links therein to FAVA, users are able to immediately find flares from the surrounding sources.
\end{itemize}

Another particularly challenging task in theoretical astrophysics is the prediction of transient phenomena (see, e.g.,~Section~\ref{sec:science_drivers}). 
Astro-COLIBRI is suitable for checking such predictions. App users receive real-time alerts for transient events as push notifications according to their specifications. This feature allows users to be notified when observations according to their model predictions occur. In~addition, all transients reported in Astro-COLIBRI can be searched via the interface by cone searches in adjustable time~windows. 

\begin{itemize}
   \item \textbf{IceCube$-$220918A and TXS 0506+056}
   Models that not only model the IceCube$-$170922A event but also predict further flares and potential neutrino detections compatible with TXS 0506+056 pose a compelling use case for Astro-COLIBRI, as~it can be facilitated to find predicted events. One example is the model of a processing AGN harbored in TXS 0506+056 \citep{Bruijn:2020ApJ...905L..13D}, which predicted the next flares around 2019--2020 and \mbox{2022--2023}. While there was no flare during the first proposed time window \endnote{Note that a flare could still be present in the offline data of IceCube.}, the~authors of~\cite{Tjus:2022} were notified through Astro-COLIBRI about a high-energy neutrino detection compatible with the location of TXS 0506+056 in the proposed period of the flare in 2022--2023. The~right mobile screen in Figure~\ref{fig:icecube_TXS} shows the cone search of IceCube-220918A and the localization of TXS 0506+056 within the uncertainty of the neutrino localization. The~customized link from Astro-COLIBRI to the TXS 0506+056 position in FAVA allowed users to immediately investigate the gamma-ray activity. Fermi-LAT gamma-ray observations \endnote{\url{https://gcn.gsfc.nasa.gov/gcn3/32565.gcn3}} of IceCube-220918A found no flare of TXS 0506+056 \endnote{Note that there is no expected correlation between the observed gamma-ray flux and the neutrino signal flux when neutrinos are produced in a gamma-absorbed environment \citep{Kun:2021}.} but detection of a new gamma-ray source, Fermi J0502.5+0037 \endnote{\url{https://www.astronomerstelegram.org/?read=15620}}. 
\end{itemize}

\section{Outlook}\label{sec:outlook}
The current landscape of highly specialized tools covering different aspects of multi-messenger astrophysics requires adaptation to the challenge of next-generation observatories through mutual connections and joint structures. The~development of Astro-COLIBRI as a holistic platform, interconnecting systems, and~services is one step in this direction. During~the first Astro-COLIBRI multi-messenger workshop in September 2022, an~agenda for the next steps of Astro-COLIBRI was defined. In~the meantime a first set of the discussed features were already implemented, such as Swift-XRT light curves. Remaining features to be implemented in the upcoming months~include:
\begin{itemize}
    \item We are developing language models with Natural Language Processing (NLP) experts to automatically {analyze} human-written messages (GCN Circulars, ATels, and~TNS AstroNotes;~\cite{alkan-wiesp2022}). The~foreseen system aims to automatically {recognize} and extract key concepts in astrophysics, such as astronomical facilities, object names, coordinates and all helpful information to trigger follow-up observation. In~perspective, we plan to deploy this system in Astro-COLIBRI to enrich the platform by performing real-time analysis of human-written reports.
    \item Tiling maps for GW uncertainty regions to facilitate follow-up observations using the algorithms described in \citep{2021JCAP...03..045A}.
    \item Support and display of ALeRCE, Astronomaly \citep{2021A&C....3600481L}, AMPEL, and~Fink alerts.
    \item More customization for push notifications, such as listening to alerts from objects, based on their visibility, etc.
    \item Technical improvements will include url-routing mechanisms, which facilitate sharing content from Astro-COLIBRI.
    \item Improved user-defined filtering criteria.
    \item Possibility to provide results of user-performed follow-up observations into Astro-COLIBRI.
\end{itemize}

Along with implementing these new features, already existing features will be maintained and improved. The~list of links to other service will be~extended.

The Astro-COLIBRI development team is welcoming comments and feedback from the community to further improve the~platform. 

{{Within} 
 the Astro-COLIBRI workshops, we encourage third-party contributions to the Astro-COLIBRI source code.} Contact: \href{mailto:astro.colibri@gmail.com}{astro.colibri@gmail.com}

\vspace{6pt} 



\authorcontributions{{Author names ordered alphabetically and not by contribution: }
Conceptualization, V.L., P.R., F.S.; methodology, V.L., P.R., F.S.; software, A.K.A., V.L., J.M., P.R., F.S.; validation, J.B.T., V.L., P.R., F.S.; formal analysis, P.R.; investigation, V.L., P.R., F.S.; resources, J.B.T., P.R., F.S.; data curation, P.R., F.S.; writing---original draft preparation, P.R.; writing---review and editing, P.R., F.S.; visualization, P.R.; supervision, J.B.T., F.S.; project administration, F.S.; funding acquisition, J.B.T., P.R., F.S. All authors have read and agreed to the published version of the manuscript.}

\funding{V.L. and F.S. acknowledge support by the European Union’s Horizon 2020 Programme under the AHEAD2020 project (grant agreement n. 871158). This work is supported by the ``ADI 2019’’ project funded by the IDEX Paris-Saclay, ANR-11-IDEX-0003-02 (P.R.). P.R. also gratefully acknowledges support by the Ruhr University Bochum Research School via the funding through the \textit{{Project} 
 International} and the \textit{Project International Event} program. J.B.T. and P.R. acknowledge funding by the BMBF, grant number 05A20PC1. J.B.T. and P.R. acknowledge the support from the \emph{Deut\-sche For\-schungs\-ge\-mein\-schaft, DFG\/} via the Collaborative Research Center SFB1491 \textit{Cosmic Interacting Matters---From Source to Signal}. P.R. acknowledges support from the German Academic Exchange Service. A.K.A., J.B.T., V.L., P.R., and F.S. acknowledge support by Institut Pascal at Université Paris-Saclay during the Paris-Saclay Astroparticle Symposiums 2021 \& 2022, with~the support of the P2IO Laboratory of Excellence (programme “Investissements d’avenir” ANR-11-IDEX-0003-01 Paris-Saclay and ANR-10-LABX-0038), the~P2I axis of the Graduate School of Physics of Université Paris-Saclay, as~well as IJCLab, CEA, IAS, IPhT, APPEC, OSUPS, the~IN2P3 master projet UCMN and EuCAPT.}

\dataavailability{The data presented in screenshots of the Astro-COLIBRI interfaces can be accessed through our publicly available API at \url{https://astro-colibri.science/} or also through the interfaces {directly}
:
\begin{itemize}
    \item Website: \url{https://astro-colibri.com} (accessed on 23 December 2022);
    \item Android App (Google Play Store): \url{https://play.google.com/store/apps/details?id=science.astro.colibri&pli=1} (accessed on 23 December 2022);
    \item iOS App (Apple App Store): \url{https://apps.apple.com/us/app/astro-colibri/id1576668763} (accessed on 23 December 2022).
\end{itemize}

Note that the exact appearance of the interface is subject to change. Here, version v2.1.0 of Astro-COLIBRI was used. {The source code of Astro-COLIBRI is currently private. Public releases are planned in the future.} {The}
 data presented in Figures~\ref{fig:1}, \ref{fig:sn} and \ref{fig:gcns} are taken from the publicly available websites/APIs \emph{voeventdb} \cite{4pisky:2016}, GCN~\cite{GCNs:2000}, and~TNS \endnote{\url{https://www.wis-tns.org/} (accessed on 23 December 2022)}, respectively. Data and the scripts used to generate these figures are available to interested researchers upon reasonable request and/or in the following repository: \url{https://github.com/reichherzerp/Transient-Statistics}.} 

\acknowledgments{{We thank the anonymous referees for insightful and stimulating~reports.} Many ideas for new features, including some prototype implementations, have been provided by the participants of the \emph{1st Astro-COLIBRI Multi-Messenger Astrophysics Workshop} \endnote{\url{https://astrophysics-workshop.web.app/} (accessed on 23 December 2022)}. We want to highlight the contributions \endnote{\url{https://indico.in2p3.fr/event/26335/sessions/16384/##20220930} (accessed on 23 December 2022)} during the Sciathon of the following partcipants:
A. Aravinthan, H. Ashkar, A. Berti, E. Blaufuss, J. Borowska, M. de Bony de Lavergne, R. Konno, M. Lincetto, I. Lypova, J. Nordin, M. Seglar-Arroyo, G. Sommani, D. Turpin, I. Viale, J. V\"olp, and S. Weimann.
We thank all partners (SFB1491, PNHE, TPIV) for their generous (financial) support and wish to express our sincere gratitude to the RUB Research School for mainly financing and supporting this international Workshop. 
We want to highlight the valuable input and feature requests from amateur astronomers. We want to also acknowledge valuable input from H.B. Eggenstein, P. Evans, F. Förster Burónand, I. Jaroschewski, E. Kun, J.P. Lenain.}

\conflictsofinterest{The authors declare no conflicts of interest. The~funders had no role in the design of the study; in the collection, analyses, or~interpretation of data; in the writing of the manuscript; or in the decision to publish the~results.} 





\begin{adjustwidth}{-\extralength}{0cm}
\printendnotes[custom] 

\reftitle{References}

\end{adjustwidth}
\end{document}